\documentstyle[12pt]{article}

  \def\pp{{\mathchoice
          {
              \kern 1pt%
              \raise 1pt
              \vbox{\hrule width5pt height0.4pt depth0pt
                    \kern -2pt
                    \hbox{\kern 2.3pt
                          \vrule width0.4pt height6pt depth0pt
                          }
                    \kern -2pt
                    \hrule width5pt height0.4pt depth0pt}%
                    \kern 1pt
           }
            {
              \kern 1pt%
              \raise 1pt
              \vbox{\hrule width4.3pt height0.4pt depth0pt
                    \kern -1.8pt
                    \hbox{\kern 1.95pt
                          \vrule width0.4pt height5.4pt depth0pt
                          }
                    \kern -1.8pt
                    \hrule width4.3pt height0.4pt depth0pt}%
                    \kern 1pt
            }
            {
              \kern 0.5pt%
              \raise 1pt
              \vbox{\hrule width4.0pt height0.3pt depth0pt
                    \kern -1.9pt  
                    \hbox{\kern 1.85pt
                          \vrule width0.3pt height5.7pt depth0pt
                          }
                    \kern -1.9pt
                    \hrule width4.0pt height0.3pt depth0pt}%
                    \kern 0.5pt
            }
            {
              \kern 0.5pt%
              \raise 1pt
              \vbox{\hrule width3.6pt height0.3pt depth0pt
                    \kern -1.5pt
                    \hbox{\kern 1.65pt
                          \vrule width0.3pt height4.5pt depth0pt
                          }
                    \kern -1.5pt
                    \hrule width3.6pt height0.3pt depth0pt}%
                    \kern 0.5pt
            }
        }}

  \def\mm{{\mathchoice
                       {
                             \kern 1pt
               \raise 1pt    \vbox{\hrule width5pt height0.4pt depth0pt
                                  \kern 2pt
                                  \hrule width5pt height0.4pt depth0pt}
                             \kern 1pt}
                       {
                            \kern 1pt
               \raise 1pt \vbox{\hrule width4.3pt height0.4pt depth0pt
                                  \kern 1.8pt
                                  \hrule width4.3pt height0.4pt depth0pt}
                             \kern 1pt}
                       {
                            \kern 0.5pt
               \raise 1pt
                            \vbox{\hrule width4.0pt height0.3pt depth0pt
                                  \kern 1.9pt
                                  \hrule width4.0pt height0.3pt depth0pt}
                            \kern 1pt}
                       {
                           \kern 0.5pt
             \raise 1pt  \vbox{\hrule width3.6pt height0.3pt depth0pt
                                  \kern 1.5pt
                                  \hrule width3.6pt height0.3pt depth0pt}
                           \kern 0.5pt}
                       }}

\catcode`@=11
\def\un#1{\relax\ifmmode\@@underline#1\else
        $\@@underline{\hbox{#1}}$\relax\fi}
\catcode`@=12

\let\du=\du                     

\def\a{\alpha}

\def\f{\phi}

\def\l{\lambda}

\def\o{\omega}

\def\t{\tau}

\def\x{\xi}
\def\z{\zeta}

\def\F{\Phi}

\def\bo{{\raise-.5ex\hbox{\large$\Box$}}}               
\def\TH{{\raise.2ex\hbox{$\displaystyle \bigodot$}\mskip-4.7mu \llap H \;}}
\def\face{{\raise.2ex\hbox{$\displaystyle \bigodot$}\mskip-2.2mu \llap {$\ddot
        \smile$}}}                                      

\def\Bar#1{\overline{#1}}                       
\def\abs#1{\left| #1\right|}                    
\def\leftrightarrowfill{$\mathsurround=0pt \mathord\leftarrow \mkern-6mu
        \cleaders\hbox{$\mkern-2mu \mathord- \mkern-2mu$}\hfill
        \mkern-6mu \mathord\rightarrow$}
\def\dvec#1{\vbox{\ialign{##\crcr
        \leftrightarrowfill\crcr\noalign{\kern-1pt\nointerlineskip}
        $\hfil\displaystyle{#1}\hfil$\crcr}}}           
\def\dt#1{{\buildrel {\hbox{\LARGE .}} \over {#1}}}     

\def\frac#1#2{{\textstyle{#1\over\vphantom2\smash{\raise.20ex
        \hbox{$\scriptstyle{#2}$}}}}}                   
\def\sfrac#1#2{{\vphantom1\smash{\lower.5ex\hbox{\small$#1$}}\over
        \vphantom1\smash{\raise.4ex\hbox{\small$#2$}}}} 
\def\bfrac#1#2{{\vphantom1\smash{\lower.5ex\hbox{$#1$}}\over
        \vphantom1\smash{\raise.3ex\hbox{$#2$}}}}       
\def\afrac#1#2{{\vphantom1\smash{\lower.5ex\hbox{$#1$}}\over#2}}    

\def\[{\lfloor{\hskip 0.35pt}\!\!\!\lceil}
\def\]{\rfloor{\hskip 0.35pt}\!\!\!\rceil}

\def\du#1#2{_{#1}{}^{#2}}

\def\ha{{\fracmm12}}

\def\un{\underline}
\def\fracmm#1#2{{{#1}\over{#2}}}

\def\low#1{{\raise -3pt\hbox{${\hskip 0.75pt}\!_{#1}$}}}

\def\Dot#1{\buildrel{_{_{\hskip 0.01in}\bullet}}\over{#1}}
\def\dt#1{\Dot{#1}}

\def\sbar#1{\stackrel{*}{\Bar{#1}}}

\newskip\humongous \humongous=0pt plus 1000pt minus 1000pt

\newif\ifdtup

\def\cqg#1#2#3{Class.~and Quantum Grav.~{\bf {#1}} (19{#2}) #3}
\def\cmp#1#2#3{Commun.~Math.~Phys.~{\bf {#1}} (19{#2}) #3}

\parskip=\medskipamount                 
\thicklines                         

\begin{document}


\thispagestyle{empty}               

\def\border{                                            
        \setlength{\unitlength}{1mm}
        \newcount\xco
        \newcount\yco
        \xco=-24
        \yco=12
        \begin{picture}(140,0)
        \put(-20,11){\tiny Institut f\"ur Theoretische Physik Universit\"at
Hannover~~ Institut f\"ur Theoretische Physik Universit\"at Hannover~~
Institut f\"ur Theoretische Physik Hannover}
        \put(-20,-241.5){\tiny Institut f\"ur Theoretische Physik Universit\"at
Hannover~~ Institut f\"ur Theoretische Physik Universit\"at Hannover~~
Institut f\"ur Theoretische Physik Hannover}
        \end{picture}
        \par\vskip-8mm}

\def\headpic{                                           
        \indent
        \setlength{\unitlength}{.8mm}
        \thinlines
        \par
        \begin{picture}(29,16)
        \put(75,16){\line(1,0){4}}
        \put(80,16){\line(1,0){4}}
        \put(85,16){\line(1,0){4}}
        \put(92,16){\line(1,0){4}}

        \put(85,0){\line(1,0){4}}
        \put(89,8){\line(1,0){3}}
        \put(92,0){\line(1,0){4}}

        \put(85,0){\line(0,1){16}}
        \put(96,0){\line(0,1){16}}
        \put(92,16){\line(1,0){4}}

        \put(85,0){\line(1,0){4}}
        \put(89,8){\line(1,0){3}}
        \put(92,0){\line(1,0){4}}

        \put(85,0){\line(0,1){16}}
        \put(96,0){\line(0,1){16}}
        \put(79,0){\line(0,1){16}}
        \put(80,0){\line(0,1){16}}
        \put(89,0){\line(0,1){16}}
        \put(92,0){\line(0,1){16}}
        \put(79,16){\oval(8,32)[bl]}
        \put(80,16){\oval(8,32)[br]}

        \end{picture}
        \par\vskip-6.5mm
        \thicklines}
  
\border\headpic {\hbox to\hsize{
\vbox{\noindent DESY 97 -- 206  \hfill October 1997 \\
ITP--UH--28/97 \hfill hep-th/9710185 }}}

\noindent
\vskip1.3cm
\begin{center}

{\Large\bf Induced scalar potentials for hypermultiplets}~\footnote{Supported 
in part by the `Deutsche Forschungsgemeinschaft'}\\
\vglue.3in

Sergei V. Ketov \footnote{
On leave of absence from:
High Current Electronics Institute of the Russian Academy of Sciences,\newline
${~~~~~}$ Siberian Branch, Akademichesky~4, Tomsk 634055, Russia}

{\it Institut f\"ur Theoretische Physik, Universit\"at Hannover}\\
{\it Appelstra\ss{}e 2, 30167 Hannover, Germany}\\
{\sl ketov@itp.uni-hannover.de}

and

Christine Unkmeir

{\it Institut f\"ur Kernphysik, Universit\"at Mainz}\\
{\it J.~J.~Becher Weg 45, 55099 Mainz, Germany}\\
{\sl unkmeir@kph.uni-mainz.de}
\end{center}
\vglue.2in

\begin{center}
{\Large\bf Abstract}
\end{center}

Charged BPS hypermultiplets can develop a non-trivial self-interaction in the 
Coulomb branch of the $N=2$ supersymmetric gauge theory, whereas neutral BPS
hypermultiplets in the Higgs branch may also have a non-trivial 
self-interaction in the presence of Fayet-Iliopoulos terms. The {\it exact\/}
hypermultiplet low-energy effective action (LEEA) takes the form of the 
non-linear sigma-model (NLSM) with a hyper-K\"ahler metric. A non-trivial 
{\it scalar} potential is also quantum-mechanically generated at non-vanishing 
central charges, either perturbatively (Coulomb branch), or non-perturbatively
(Higgs branch). We calculate the effective scalar potentials for (i) a single 
charged hypermultiplet in the Coulomb branch and (ii) a single neutral 
hypermultiplet in the Higgs branch. The first case corresponds to the NLSM with
the Taub-NUT (or KK-monopole) metric for the kinetic LEEA, whereas the second 
one is attached to the NLSM having the Eguchi-Hanson instanton metric.
 
\newpage

\section{Introduction}

The $N=1$ supersymmetric matter in four spacetime dimensions is described in 
terms of chiral $N=1$ multiplets and linear $N=1$ multiplets, that are 
(field-theory) dual to each other. The $N=1$ chiral superfields $\F$ may have 
a chiral scalar superpotential described by a holomorphic function $W(\F)$. As 
regards the fundamental quantum field theory actions, the function $W(\F)$ 
should be restricted to a cubic polynomial by renormalizability, while there 
is no such restriction if it appears in the low-energy effective action (LEEA)
of a quantum $N=1$ supersymmetric field theory. In the exact LEEA, the full
quantum-generated scalar potential $W(\F)$ is supposed to include all 
perturbative as well as all non-perturbative corrections, if any.

The $N=2$ supersymmetric matter is described by hypermultiplets. The off-shell
hypermultiplets come in two fundamental versions that are (field-theory) dual
to each other in the $N=2$ harmonic superspace.~\footnote{See ref.~\cite{all}
for a recent review and an introduction to the harmonic superspace.} The first
version is called a Fayet-Sohnius-type (FS) hypermultiplet, it is described by
an unconstrained complex analytic superfield $q^+$ of $U(1)$-charge $(+1)$, and
it is on-shell equivalent to the standard Fayet-Sohnius hypermultiplet 
comprising two $N=1$ chiral supermultiplets. The second version is called a
Howe-Stelle-Townsend-type (HST) hypermultiplet, it is described by a real
unconstrained analytic superfield $\o$ of vanishing $U(1)$-charge, and it is
on-shell equivalent to the standard $N=2$ tensor (or $N=2$ linear) multiplet
comprising an $N=1$ chiral multiplet and an $N=1$ linear multiplet. Unlike that
in $N=1$ supersymmetry, there is apparently no $N=2$ supersymmetric invariant 
that would generate the scalar potential for the scalar components of a
hypermultiplet. It is often assumed in the literature that the hypermultiplet
scalar potential (beyond the BPS mass term generated by central charges) 
simply does not exist, both in a fundamental $N=2$ supersymmetric field theory
action {\it and\/} in the corresponding LEEA provided that $N=2$ supersymmetry 
is not broken. 

At the fundamental level, any non-trival hypermultiplet potential is indeed
forbidden by renormalizability and unitarity. However, contrary to the naive 
expectations, a non-trivial scalar potential does appear in the hypermultiplet
 LEEA provided that the central charges do not vanish. It
was recently demonstrated~\cite{ikz} in the case of a single charged (FS-type)
hypermultiplet minimally coupled to an abelian $N=2$ vector multiplet (i.e.
in the Coulomb branch). In this Letter we give some more details of this
calculation (sect.~2), and then extend it to yet another interesting case of a 
single (HST-type) hypermultiplet with a Fayet-Iliopoulos (FI) term (sect.~3).
\vglue.2in

\section{Taub-NUT action with central charges}

A FS-type hypermultiplet is most naturally described in the $N=2$ harmonic
superspace, in terms of an unconstrained complex analytic superfield $q^+$ of
$U(1)$-charge $(+1)$. As was shown in ref.~\cite{ikz}, a single FS-type charged
hypermultiplet with a non-vanishing central charge (or BPS mass) gets the 
one-loop induced self-interaction in the Coulomb branch of an $N=2$ 
supersymmetric gauge theory. The hypermultiplet LEEA is given by the NLSM 
whose target space metric is Taub-NUT.

The corresponding $N=2$ harmonic superspace action in the analytic subspace 
$\z^M=(x^m_A,{\theta^+}\low{\a},\bar{\theta}^+_{\dt{\a}})$ 
reads as follows:~\footnote{We use the standard notation for the $N=2$ 
harmonic superspace, see e.g., ref.~\cite{all}.}
\begin{eqnarray}
S_T[q]= -\int d\z^{(-4)} du \left\{~ \sbar{q}{}^+D^{++}q^+ 
+\fracmm{\lambda}{2}(q^+)^2(\sbar{q}{}^+)^2 ~\right\}
\label{action}
\end{eqnarray}
where the covariant derivative $D^{++}$, in the analytic basis at 
non-vanishing central charges $Z$ and $\Bar{Z}$, has been introduced,
\begin{eqnarray}
D^{++} =\partial^{++}
-2i\theta^+\sigma^m\bar{\theta}^+\partial_m +i\theta^+\bar{\theta}^+\Bar{Z}
+i\bar{\theta}{}^+\bar{\theta}{}^+Z
\label{der}
\end{eqnarray}
and $\l$ is the induced (Taub-NUT) NLSM coupling constant. Eq.~(\ref{der})
can be most easily obtained by (Scherk-Schwarz) dimensional reduction from six
dimensions~\cite{ikz}. For simplicity, we ignore here possible couplings to an
abelian $N=2$ vector superfield. The explicit expression for $\l$ in terms of 
the fundamental gauge coupling and the hypermultiplet BPS mass can be found in
ref.~\cite{ikz}. Our $q^+$ superfields are of dimension minus one (in units of 
length), while the coupling constant $\l$ is of dimension two.~\footnote{We
assume that $c=\hbar=1$ as usual.} 

Our aim now is to find the component form of the action (\ref{action}). 
Without central charges it was done in ref.~\cite{gios}. The non-vanishing 
central charges were incorporated in ref.~\cite{ikz}. In this section, we are 
going to concentrate on the component scalar potential originating from the
action (\ref{action}). The corresponding equations of motion are
\begin{equation}
D^{++}q^+ +\lambda(q^+\sbar{q}{}^+)q^+ = 0 \quad {\rm and}\quad 
D^{++}\sbar{q}{}^+ -\lambda(q^+\sbar{q}{}^+)\sbar{q}{}^+ = 0 
\label{eqmo}
\end{equation}

Since we are only interested in the purely bosonic part of the action 
(\ref{action}), we drop all fermionic fields in the $\theta^+$,
$\bar{\theta^+}$ expansion of $q^+$,
\begin{eqnarray}
q^+(\z,u) &=& F^+(x_A,u) + i\theta^+\sigma^m\bar{\theta}^+A^-_m(x_A,u) 
\nonumber\\
& &+\theta^+\theta^+M^-(x_A,u)+\bar{\theta}^+\bar{\theta}^+N^-(x_A,u)
\nonumber\\
& & +\theta^+\theta^+\bar{\theta}^+\bar{\theta}^+P^{(-3)}(x_A,u) 
\label{exp}
\end{eqnarray}
Being substituted into eq.~(\ref{eqmo}), eq.~(\ref{exp}) yields
\begin{eqnarray}
\label{bw1}
\partial^{++}F^+ +\lambda(F^+\sbar{F}{}^+)F^+ &=& 0 \\
\label{bw2}
\partial^{++}A^-_m - 2\partial_mF^+ +2\lambda A^-_m \sbar{F}{}^+F^+ 
+ \lambda (F^+)^2 \sbar{A}_m{}^- &=& 0 \\
\label{bw3}
\partial^{++}M^- +2\lambda M^- \sbar{F}{}^+F^+ + \lambda \sbar{N}{}^-(F^+)^2 
+i\Bar{Z}F^+ &=& 0 \\
\label{bw4}
\partial^{++}N^- +2\lambda N^- \sbar{F}{}^+F^+ + \lambda \sbar{M}{}^-(F^+)^2 
+iZF^+ &=& 0 \\
\label{bw5}
\partial^{++} P^{(-3)} + \partial^m A^-_m + 2\lambda F^+\sbar{F}{}^+P^{(-3)}
+\lambda(F^+)^2\sbar{P}{}^{(-3)} & & \nonumber \\
-\fracmm{\lambda}{2} A^{-m} A^-_m \sbar{F}{}^+ 
-\lambda A^{-m}\sbar{A}_m{}^- F^+
+2\lambda F^+(M^-\sbar{M}{}^- +N^- \sbar{N}{}^-) & &\nonumber \\
 +2\lambda \sbar{F}{}^+M^-N^- +i\Bar{Z}N^- +iZM^- &=& 0 
\end{eqnarray}
as well as their conjugates.

The integration over $\theta^+, \bar{\theta}^+$ in eq.~(\ref{action}) results 
in the bosonic action
\begin{eqnarray}
S_T &=& 
-\int d\z^{(-4)} du \left\{~ 
\sbar{q}{}^+D^{++}q^+ +\fracmm{\lambda}{2}(q^+)^2
(\sbar{q}{}^{+})^2 ~\right\}\nonumber\\
&\to & S= -\frac{1}{2}\int d^4xdu~ \left\{~ 
(\sbar{A}_m{}^-\partial^mF^+ - A^-_m\partial^m\sbar{F}{}^+)\right.\nonumber\\
& &\left. 
+iF^+(\Bar{Z}\sbar{M}{}^- +Z\sbar{N}{}^-)+i \sbar{F}{}^+(ZM^-+\Bar{Z}N^-)
~\right\}
\label{baction}
\end{eqnarray}
Since the action (\ref{action}) has the global $U(1)$ invariance
\begin{eqnarray}
q^{+~'}=e^{i\alpha}q^+, \qquad \sbar{q}{}^{+'}=e^{-i\alpha}\sbar{q}{}^+
\end{eqnarray}
there exists the conserved Noether current $j^{++}$
\begin{eqnarray}
D^{++}j^{++}=0, \qquad j^{++}=iq^+\sbar{q}{}^+ 
\end{eqnarray}
It implies, in particular, that $\partial^{++}(F^+\sbar{F}{}^+)=0$ and, hence,
\begin{eqnarray}
F^+(x,u)\sbar{F}{}^+(x,u)=C^{(ij)}u^+_{~i}u^+_{~j} \\
(F^+\sbar{F}{}^+)^{\frac{*}{}} = -F^+\sbar{F}{}^+ \rightarrow 
\Bar{C^{(ij)}}= -\epsilon_{il}\epsilon_{jn}C^{(ln)}~,
\end{eqnarray}
where the new function $C^{(ij)}(x)$ has been introduced. Changing the 
variables as
\begin{eqnarray}
F^+(x,u)=f^+(x,u)e^{\lambda \varphi}, \qquad 
\varphi(x,u)=-C^{(ij)}(x)u^+_{~i}u^-_{~j}=-\sbar{\varphi}(x,u)
\end{eqnarray}
now reduces eq.~(\ref{bw1}) to the linear equation
\begin{eqnarray}
\partial^{++}f^+(x,u) = 0 \rightarrow f^+(x,u)= f^i(x)u^+_{~i}
\end{eqnarray}
After taking into account that
\begin{eqnarray}
F^+\sbar{F}{}^+=f^+\sbar{f}{}^+ \rightarrow 
C^{(ij)}(x)=-f^{(i}(x)\bar{f}^{j)}(x)~,
\end{eqnarray}
where $\bar{f}^i = \epsilon^{ij}\bar{f}_j$ and $\bar{f}_j\equiv\Bar{(f^j)}$,
we obtain a general solution in the form
\begin{eqnarray}
F^+(x,u)&=&f^{i}u_{~i}^{+} e^{\lambda\varphi} \nonumber \\
&=& f^{i}(x)u^+_{~i}\exp\{ \lambda f^{(j}\bar{f}^{k)}u_{~j}^{+}u_{~k}^{-} \}
\end{eqnarray}
The same conclusion also appears when using the {\it Ansatz}
\begin{eqnarray}
F^{+}= e^{C}[ f^{i}u_{~i}^{+}+ B^{ijk}u_{i}^{+}u_{~j}^{+}u_{~k}^{-} ]
\label{1ansatz}
\end{eqnarray}
in terms of functions $C$ and $B^{ijk}$ at our disposal. After 
substituting eq.~(\ref{1ansatz}) into the equation of motion (\ref{bw1}),
we find
\begin{equation}
B^{ijk}= 0  \qquad {\rm and} \qquad
C = \lambda f^{(i}\bar{f}^{j)}u_{~i}^{+}u_{~j}^{-}
\end{equation}
so that 
\begin{eqnarray}
F^{+}= f^{i}u_{i}^{+}\exp\{ \lambda f^{(j}\bar{f}^{k)}u_{~j}^{+}u_{~k}^{-} \}
\end{eqnarray}
again. To get a similar equation for $A^{-}_{m}$, we use eq.~(\ref{bw2})
and the {\it Ansatz}
\begin{eqnarray}
A^{-}_{m} &=& e^{\lambda \varphi} \left\{ a\lambda f^{i}u_{~i}^{+}
\partial_{m}(f^{(k}\bar{f}^{j)}u^{-}_{~k}u^{-}_{~j}) \right. \nonumber\\
& &\left. +b \partial_{m}f^{i}u_{~i}^{-}
+c\, \fracmm{\lambda f^{i}u_{~i}^{-}}{1+\lambda f\bar{f}}
(f^{j}\partial_{m}\bar{f}_{j}-\bar{f}_{j}\partial_{m}f^{j})
\right\}
\label{2ansatz}
\end{eqnarray}
with some coefficients $(a,b,c)$ to be determined. After substituting 
eq.~(\ref{2ansatz}) into eq.~(\ref{bw2}), we find the relations
\begin{equation}
b = 2~, \quad 2a=2+b~, \quad{\rm and}\quad c= b-a+1 
\end{equation}
so that $a = b = 2$ and $c = 1$. Therefore, we have
\begin{eqnarray}
A^{-}_{m} &=& e^{\lambda \varphi} \left\{ 2\lambda f^{i}u_{~i}^{+}
\partial_{m}(f^{(k}\bar{f}^{j)}u^{-}_{~k}u^{-}_{~j})\right. \\ \nonumber
& & \left. +2 \partial_{m}f^{i}u_{~i}^{-}
+ \fracmm{\lambda f^{i}u_{~i}^{-}}{1+\lambda f\bar{f}}
(f^{j}\partial_{m}\bar{f}_{j}-\bar{f}_{j}\partial_{m}f^{j})
\right\}
\end{eqnarray}

To solve the remaining equations of motion (\ref{bw3}) and (\ref{bw4})
for the auxiliary fields $M^{-}$ and $N^{-}$ (the rest of equations of motion 
in eq.~(\ref{bw5}) is irrelevant for our purposes), 
we introduce the {\it Ansatz}
\begin{eqnarray}
\label{m}
M^{-}&=&e^{\lambda \varphi}R^{-}\equiv R e^{\lambda\varphi}f^{i}u^{-}_{~i}\\
\label{n}
N^{-}&=&e^{\lambda \varphi}S^{-}\equiv S e^{\lambda \varphi}f^{i}u^{-}_{~i}
\end{eqnarray}
with some coefficient functions $R$ and $S$ to be determined.
After substituting eq.~(\ref{m}) into eq.~(\ref{bw3}) we get
\begin{eqnarray*}
\partial^{++}R^{-}
-\lambda f^{(j}\bar{f}^{k)}u^{+}_{~j}u^{+}_{~k}R^{-}
+\lambda f^{i}f^{j}u^{+}_{~i}u^{+}_{~j}\sbar{S}{}^{-}
+i\Bar{Z}f^{i}u^{+}_{~i}&=&0 \nonumber \\
Rf^{i}u^{+}_{~i}
-R\lambda f^{(m}\bar{f}^{n)}f^{i}u^{+}_{~m}u^{+}_{~n}u^{-}_{~i}
+i\Bar{Z}f^{i}u^{+}_{~i} 
-\lambda\bar{S}f^{m}f^{n}\bar{f}^{i}u^{+}_mu^{+}_{~n}u^{-}_{~i}&=& 0 
\nonumber \\
Rf^{i}u^{+}_{~i}
+R\lambda f^{m}\bar{f}_{n}f^{i}(u^{-~n}u^{+}_{~i}+\delta^{n}_{i})u^+_m
+i\Bar{Z}f^{i}u^{+}_{~i}
-\lambda\bar{S}f^{m}f^{n}\bar{f}^{i}u^{+}_mu^{+}_{~n}u^{-}_{~i}&=& 0 
\nonumber\\
f^{i}u^{+}_{~i}[R(1+\lambda f^{j}\bar{f}_{j})+i\Bar{Z}]
-\lambda(R+\bar{S})f^{m}\bar{f}^{n}f^{i}u^{+}_{~m}u^{+}_{~i}u^{-}_{~n}&=& 0
\nonumber \\
\end{eqnarray*}
It follows
\begin{equation}
R=-\,\fracmm{i\Bar{Z}}{1+\lambda f\bar{f}} \quad {\rm and,~hence,}
\quad
M^{-}=-\,\fracmm{i\Bar{Z}}{1+\lambda f\bar{f}}\,e^{\lambda \varphi}
f^{i}u^{-}_{~i}
\end{equation}

Similarly, we find from eqs.~(\ref{bw4}) and (\ref{n}) that
\begin{eqnarray}
N^{-}= -\,\fracmm{iZ}{1+\lambda f\bar{f}}\,e^{\lambda \varphi}f^{i}u^{-}_{~i}
\end{eqnarray}
Substituting now the obtained solutions for the auxiliary fields 
$F^{+},A^{-}_{a},M^{-}$ and $N^{-}$ into the action (\ref{baction}) yields the
bosonic NLSM action  
\begin{equation}
S = \frac{1}{2}\int d^{4}x\, \left\{ 
g_{ij}\partial_{m}f^{i}\partial^{m}f^{j}
+\bar{g}^{ij}\partial_{m}\bar{f}_{i}\partial^{m}\bar{f}_{j}
+2h^{i}_{~j}\partial_{m}f^{j}\partial^{m}\bar{f}_{i}) -V(f)\right\}
\end{equation}
whose metric takes the form~\cite{gios}
\begin{eqnarray}
g_{ij}&=& \fracmm{\lambda(2+\lambda f\bar{f})}{2(1+\lambda f\bar{f})}
\bar{f}_i\bar{f}_j \nonumber\\
\bar{g}^{ij}&=&  \fracmm{\lambda(2+\lambda f\bar{f})}{2(1+\lambda f\bar{f})}
f^if^j \\
h^i_j&=& \delta^i_j(1+\lambda f\bar{f})  
-\fracmm{\lambda(2+\lambda f\bar{f})}{2(1+\lambda f\bar{f})}f^i\bar{f}_j
\nonumber
\label{1nlsm}
\end{eqnarray}
This metric is known to be equivalent to the standard Taub-NUT metric up to a 
field redefinition~\cite{gios}. The scalar potential in eq.~(\ref{1nlsm}) takes
the form~\cite{ikz}
\begin{equation}
V(f)=\fracmm{Z\Bar{Z}}{1+\lambda f\bar{f}}\,f\bar{f}
\label{taubpot}
\end{equation}

By construction, the effective scalar potential (\ref{taubpot}) for a single
charged hypermultiplet is generated in the one-loop perturbation 
theory~\cite{ikz}, and it is exact in the Coulomb branch. The vacuum 
expectation  values for the scalar hypermultiplet components, which are to be
calculated from this effective potential, all vanish. Notably, the BPS mass 
$m^2_{\rm BPS}=\abs{Z}^2$ is not renormalized, as it should. Nevertheless,
the whole effective scalar potential (\ref{taubpot}) is not merely the 
quadratic (BPS mass) contribution.
\vglue.2in

\section{Eguchi-Hanson action with central charges}

As was argued in refs.~\cite{all,ikz}, a non-trivial hypermultiplet 
self-interaction can be non-perturbatively generated in the Higgs branch, in the
presence of non-vanishing constant FI-term 
$\x^{(ij)}=\ha(\vec{\t}\cdot\vec{\x})^{ij}$, 
where $\vec{\t}$ are Pauli matrices. The FI-term is nothing but the 
vacuum expectation value of the $N=2$ vector multiplet auxiliary components 
(in a WZ-like gauge). The FI-term has a nice geometrical interpretation in the
underlying ten-dimensional type-IIA superstring brane picture made out of two 
solitonic 5-branes located at particular values of $\vec{w}=(x^7,x^8,x^9)$ and
some Dirichlet 4- and 6-branes, all having the four-dimensional spacetime 
$(x^0,x^1,x^2,x^3)$ as the common macroscopic world-volume~\cite{witten}. The 
values of $\vec{\x}$ can then be identified with the {\it angles\/} at which 
the two 5-branes intersect, $\vec{\x}=\vec{w}_1-\vec{w}_2$, in the type-IIA 
picture~\cite{all}. The three hidden dimensions $(\vec{w})$ are identified
by the requirements that they do not include the two hidden dimensions 
$(x^4,x^5)$ already used to generate central charges in the effective 
four-dimensional field theory, and that they are to be orthogonal to the 
direction $(x^6)$ in which the Dirichlet 4-branes are finite and terminate on 
5-branes.

The simplest non-trivial LEEA for a single dimensionless $\o$-hypermultiplet 
in the Higgs branch reads~\cite{all,ikz}:
\begin{equation}
S_{EH}[\o]=-\,\fracmm{1}{2\kappa^2}\int d\zeta^{(-4)} du \left\{ 
\left(D^{++}\omega\right)^2
-\fracmm{(\x^{++})^2}{\omega^2}\right\} 
\label{omega}
\end{equation}
where $\x^{++}=u^+_iu^+_j\x^{(ij)}$ is the FI-term, and $\kappa$ is the
coupling constant of dimension one (in units of length). When changing the
variables to $q^+_a=u^+_a\omega + u^-_af^{++}$ and eliminating the Lagrange 
multiplier $f^{++}$ via its algebraic equation of motion, one can rewrite 
eq.~(\ref{omega}) to an equivalent form in terms of a 
$q^+$-hypermultiplet as follows~\cite{giot}: 
\begin{equation}
S_{EH}[q]=-\,\fracmm{1}{2\kappa^2}
\int d\zeta^{(-4)} du \left\{  q^{a+} D^{++}q^+_{a} -
\fracmm{(\x^{++})^2}{(q^{a+}u^-_{a})^2}\right\} 
\label{qu}
\end{equation}
where we have used the notation $q_a^+=(\sbar{q}{}^+,q^+)$ and $q^{a+}=
\varepsilon^{ab}q^+_b$. In its turn, eq.~(\ref{qu}) is classically equivalent
to the following gauge-invariant action in terms of {\it two\/} FS 
hypermultiplets $q^+_{aA}$ $(A=1,2)$ and the auxiliary real analytic $N=2$ 
vector 
superfield $V^{++}$~\cite{giot}:
\begin{equation}
S_{EH}[q,V]=-\,\fracmm{1}{2\kappa^2}
\int d\zeta^{(-4)} du \left\{ q^{a+}_A D^{++}q^+_{aA}+
V^{++}\left(\frac{1}{2}\varepsilon^{AB}q^{a+}_Aq^+_{Ba}+\x^{++}\right)
\right\}
\label{qv}
\end{equation}
We now calculate the component form of this hypermultiplet self-interaction
by using eq.~(\ref{qv}) as our starting point. In a bit more explicit form, it 
reads  
\begin{eqnarray}
S =-\fracmm{1}{2\kappa^2}
\int d\z^{(-4)} du \left\{ \sbar{q}_{1}{}^{+}D^{++}q_{1}^{+}
+\sbar{q}_{2}{}^{+}D^{++}q_{2}^{+}+
V^{++}(\sbar{q}_{1}{}^{+}q_{2}^{+}-\sbar{q}_{2}{}^{+}q_{1}^{+}+\xi^{++})
\right\}
\label{ehaction}
\end{eqnarray}

The equations of motion are given by
\begin{eqnarray}
\label{bew1}
D^{++}q_{1}^{+}+V^{++}q_{2}^{+}&=& 0 \\
\label{bew2}
D^{++}q_{2}^{+}-V^{++}q_{1}^{+}&=& 0 \\
\label{bew3}
\sbar{q}_{1}{}^{+}q_{1}^{+}-\sbar{q}_{2}{}^{+}q_{1}^{+}+\xi^{++}&=&0
\end{eqnarray}
while the last equation is clearly the algebraic constraint on the two FS
hypermultiplets. In what follows, we ignore fermionic contributions and use a
WZ-gauge for the $N=2$ vector superfield $V^{++}$, so that $D^{++}$ and $q^+$
are still given by eqs.~(\ref{der}) and (\ref{exp}), whereas
\begin{eqnarray}
V^{++}&=& -2i\theta^{+}\sigma^{m}\bar{\theta}^{+}V_{m}(x_{A})
+\theta^+\theta^+\bar{a}(x_{A})+\bar{\theta}^+\bar{\theta}^+a(x_{A})\\
& & +\theta^+\theta^+\bar{\theta}^+\bar{\theta}^+ 
D^{(ij)}(x_A)u^{-}_{~i}u^{-}_{~j}
\end{eqnarray}
The equation of motion (\ref{bew1}) in components reads
\begin{eqnarray}
\label{eq11}
\partial^{++}F_{1}^{+}&=& 0 \\
\label{eq12}
-2\partial_{m}F^{+}_{1}+\partial^{++}A^{-}_{1m}-2V_{m}F^{+}_{2} &=& 0 \\
\label{eq13}
i\Bar{Z}F^{+}_{1}+\partial^{++}M^{-}_{1}+\bar{a}F^{+}_{2}&=&0 \\
\label{eq14}
iZF^{+}_{1}+\partial^{++}N^{-}_{1}+aF^{+}_{2} &=& 0 \\
\label{eq15}
\partial^{++}P^{(-3)}_{1}+\partial^{m}A^{-}_{1m}+i\bar{Z}N^{-}_{1}+iZM^{-}_{1}
& & \nonumber \\
+V^{m}A^{-}_{2m}+\bar{a}N^{-}_{2}+aM^{-}_{2}
+D^{(ij)}u^{-}_{~i}u^{-}_{~j}F^{+}_{2} &=&0
\end{eqnarray}
whereas eq.~(\ref{bew2}) gives
\begin{eqnarray}
\label{eq21}
\partial^{++}F_{2}^{+}&=& 0\\
\label{eq22}
-2\partial_{m}F^{+}_{2}+\partial^{++}A^{-}_{2m}+2V_{m}F^{+}_{1} &=& 0 \\
\label{eq23}
i\bar{Z}F^{+}_{2}+\partial^{++}M^{-}_{2}-\bar{a}F^{+}_{1}&=&0 \\
\label{eq24}
iZF^{+}_{2}+\partial^{++}N^{-}_{2}-aF^{+}_{1} &=& 0 \\
\label{eq25}
\partial^{++}P^{(-3)}_{2}+\partial^{m}A^{-}_{2m}
+i\bar{Z}N^{-}_{2}+iZM^{-}_{2} & & \nonumber \\
-V^{m}A^{-}_{1m}-\bar{a}N^{-}_{1}-aM^{-}_{1}
-D^{(ij)}u^{-}_{~i}u^{-}_{~j}F^{+}_{1}&=&0 
\end{eqnarray}
The constraint (\ref{bew3}) in components is given by
\begin{eqnarray}
\label{eq31}
\sbar{F}_1{}^{+}F^{+}_{2}-\sbar{F}_2{}^{+}F^{+}_{1}+\xi^{++}&=& 0 \\
\label{eq32}
\sbar{A}_{1a}{}^- F^{+}_{2}+\sbar{F}_1{}^{+}A^{-}_{2a}
-\sbar{A}_{2a}{}^- F^{+}_{1}-\sbar{F}_2{}^{+}A^{-}_{1a}&=& 0 \\
\label{eq33}
\sbar{F}_1{}^{+}M^{-}_{2}-\sbar{F}_2{}^{+}M^{-}_{1}+\sbar{N}_1{}^{-}F^{+}_{2}
-\sbar{N}_2{}^{-}F^{+}_{1} &=& 0 \\
\label{eq34}
\sbar{F}_1{}^{+} N^{-}_{2}-\sbar{F}_2{}^{+}N^{-}_{1}+\sbar{M}_1{}^{-}F^{+}_{2}
-\sbar{M}_2{}^{-}F^{+}_{1} &=& 0  \\
\label{eq35}
-\frac{1}{2}\sbar{A}_1{}^{m-}A^{-}_{2m}+\sbar{M}_1{}^{-}M^{-}_{2}
+\sbar{N}_1{}^{-}N^{-}_{2}+\sbar{P}_1{}^{(-3)} F^{+}_{2}
+\sbar{F}_1{}^{+} P^{(-3)}_{2} & & \nonumber\\
+\frac{1}{2}\sbar{A}_2{}^{m-}A^{-}_{1m}-\sbar{M}_2{}^{-} M^{-}_{1}
-\sbar{N}_2{}^{-} N^{-}_{1}-\sbar{P}_2{}^{(-3)} F^{+}_{1}
-\sbar{F}_2{}^{+} P^{(-3)}_{1} &=& 0 \nonumber \\
\end{eqnarray}
Substituting the component expressions for $q_{A}^{+}$ and $V^{++}$ into 
the action (\ref{ehaction}) results in the following bosonic action:
\begin{eqnarray}
S &=& -\,\fracmm{1}{2\kappa^2}\int d^{4}x du\left\{
\sbar{F}_1{}^{+}\partial^{m}A^{-}_{1m}+\sbar{F}_2{}^{+}\partial^{m}A^{-}_{2m}
+V^{m}(\sbar{F}_1{}^{+}A^{-}_{2m}-\sbar{F}_2{}^{+}A^{-}_{1m})\right.\nonumber\\
& & +\sbar{a}(\sbar{F}_1{}^{+}N^{-}_{2}-\sbar{F}_2{}^{+} N^{-}_{1})
+a (\sbar{F}_1{}^{+}M^{-}_{2}-\sbar{F}_2{}^{+} M^{-}_{1})\nonumber\\
& & +iD^{(ij)}u^{-}_{~i}u^{-}_{~j}(\x^{++}+\sbar{F}_1{}^+F^+_2
-\sbar{F}_2{}^+ F^+_1)\nonumber\\
& &\left.  +\sbar{F}_1{}^{+}(i\Bar{Z}N^{-}_{1}+iZM^{-}_{1})
+\sbar{F}_2{}^{+}(i\Bar{Z}N^{-}_{2}+iZM^{-}_{2})\right\}
\label{egaction}
\end{eqnarray}

The next step in our calculation is to fix the harmonic dependence of the 
fields $F^{+}_{i},A^{-}_{ia},M^{-}_{i}$ and $N^{-}_{i}$. Eqs.~(\ref{eq11}) 
and (\ref{eq21}) imply
\begin{eqnarray}
F^{+}_{1} = f^{i}_{1}u^{+}_{~i} \quad{\rm and}\quad  
F^{+}_{2} = f^{i}_{2}u^{+}_{~i}
\label{Ff}
\end{eqnarray}
whereas eq.~(\ref{eq12}) yields
\begin{eqnarray}
-2\partial_{m}F^{+}_{1}+\partial^{++}A^{-}_{1m}-2V_{m}F^{+}_{2} = 0 
\end{eqnarray}
After introducing the {\it Ansatz}
\begin{eqnarray}
A^{-}_{1m}= A_{1m}^{i}u^{-}_{~i}+B_{1m}^{ijk}u^{+}_{~i}u^{-}_{~j}u^{-}_{~k}
\end{eqnarray}
we find that
\begin{eqnarray}
A^{-}_{1m}= (2 \partial_{m}f^{i}_{1}+2V_{m}f^{i}_{2})u^{-}_{~i}
\label{A1f}
\end{eqnarray}
Similarly, it follows from  eq.~(\ref{eq22}) that
\begin{eqnarray}
A^{-}_{2m}= (2 \partial_{m}f^{i}_{2}-2V_{m}f^{i}_{1})u^{-}_{~i}
\label{A2f}
\end{eqnarray}
Eqs.~(\ref{eq13}), (\ref{eq14}), (\ref{eq23}) and (\ref{eq24}) now imply
\begin{eqnarray}
\label{M1f}
M^{-}_{1} &=& -(\sbar{a}f^{i}_{2}+i\Bar{Z}f^{i}_{1})u^{-}_{~i} \\
\label{N1f}
N^{-}_{1} &=& -(af^{i}_{2}+iZf^{i}_{1})u^{-}_{~i} \\
\label{M2f}
M^{-}_{2} &=& (\sbar{a}f^{i}_{1}-i\Bar{Z}f^{i}_{2})u^{-}_{~i} \\
\label{N2f}
N^{-}_{2} &=& (af^{i}_{1}-iZf^{i}_{2})u^{-}_{~i} 
\end{eqnarray}
After substituting all the component solutions into the action 
(\ref{egaction}), we find the (abelian) gauged NLSM action
\begin{eqnarray}
S &=&\fracmm{1}{2\kappa^2}\int d^{4}x\left\{
(\partial_{m}f_{1}^{~i}+V_{m}f_{2}^{~i})
(\partial^{m}\bar{f}_{1i}+V^{m}\bar{f}_{2i})
+(\partial_{m}f_{2}^{~i}-V_{m}f_{1}^{~i})
(\partial^{m}\bar{f}_{2i}-V^{m}\bar{f}_{1i}) \right. \nonumber\\
& &\left. -\,\fracmm{Z\Bar{Z}}{(f_1\bar{f}_1+f_2\bar{f}_2)}
\left[(f_1^i\bar{f}_{2i}-f_2^i\bar{f}_{1i})^2
+(f_1^i\bar{f}_{1i}+f_2^i\bar{f}_{2i})^2\right]\right\} \nonumber\\
\label{wirkung2}
\end{eqnarray}
where the scalar hypermultiplet components $f^i_{1,2}$ are still subject to the
constraint 
\begin{eqnarray}
\x^{(ij)}&=& \bar{f}_{1}^{(i}f^{j)}_2-f^{(i}_1\bar{f}^{j)}_2 
\label{xi}
\end{eqnarray}
In calculating the action (\ref{wirkung2}) we have also used the equation
of motion for the $N=2$ vector multiplet auxiliary field $a$, whose solution
reads
\begin{eqnarray}
\label{a}
a&=& -iZ\fracmm{f_1^i\bar{f}_{2i}-f_2^{i}\bar{f}_{1i}}
{f_1^i\bar{f}_{1i}+f_2^i\bar{f}_{2i}}
\end{eqnarray}
A solution to the equation of motion for the vector gauge field $V_m$ is given
by
\begin{eqnarray}
2V_m &=&\fracmm{\partial_m\bar{f}_{1j}f_2^j
-\bar{f}_{1j}\partial_mf_2^j-\partial_m\bar{f}_{2j}f_1^j
+\bar{f}_{2j}\partial_m f_1^j}
{\bar{f}_{1j}f_1^j+\bar{f}_{2j}f_2^j}
\label{va}
\end{eqnarray}

In terms of the two complex scalar $SU(2)$ doublets $f^i_{1,2}$ subject to the
three real constraints (\ref{xi}) and one abelian gauge invariance, we have
$2\times 2\times 2 - 3 -1=4$ independent degrees of freedom, as it should. 
After eliminating the auxiliary vector potential $V_m$ via eq.~(\ref{va}) and 
solving the constraint (\ref{xi}), one finds the NLSM with a non-trival 
hyper-K\"ahler metric (by construction, as the consequence of $N=2$ 
supersymmetry) {\it and\/} a non-trivial scalar potential
\begin{equation}
V=\fracmm{Z\Bar{Z}}{(f_1\bar{f}_1+f_2\bar{f}_2)}
\left[ (f_1^i\bar{f}_{2i}-f_2^i\bar{f}_{1i})^2
+(f_1^i\bar{f}_{1i}+f_2^i\bar{f}_{2i})^2\right]
\label{pot}
\end{equation}
The kinetic terms in the NLSM (\ref{wirkung2}) are known to represent the 
{\it Eguchi-Hanson} instanton metric up to a field redefinition~\cite{giot}, 
so that to this end we concentrate on the scalar potential (\ref{pot})
only. 

Let's introduce the following notation
\begin{eqnarray}
\bar{f}_{(1,2)}^{~1} = \stackrel{\ast}{f}_{(1,2)}{}^{2}~~, \qquad 
\bar{f}_{(1,2)}^{~2} = -\stackrel{\ast}{f}_{(1,2)}{}^{1}
\label{nota}
\end{eqnarray}
and keep the positions of indices as above. The operator $\ast$ denotes 
the usual complex conjugation. The constraints (\ref{xi}) now take the form
\begin{eqnarray*}
\x^{11}&=& \bar{f}_{1}^{1}f_2^{~1}-f_1^{~1}\bar{f}_2^{~1}=
\stackrel{\ast}{f}_1{}^{2}f_2^{~1}- f_1^{~1}\stackrel{\ast}{f}_2{}^{2}
\nonumber \\
\x^{12}&=& \frac{1}{2}(\bar{f}_1^{~1}f_2^{~2}+\bar{f}_1^{~2}f_2^{~1})
-\frac{1}{2}(f_1^{~1}\bar{f}_2^{~2}+f_1^{~2}\bar{f}_2^{~1}) \nonumber\\ 
\x^{22}&=& \bar{f}_{1}^{2}f^2_2-f^2_1\bar{f}^2_2 \nonumber\\ 
\end{eqnarray*}
When mulitplying these constraints with Pauli matrices 
$(\tau_1, 1,\tau_3)_{ij}$, we get 
\begin{eqnarray}
\x^1&=& \bar{f}_{1}^{1}f_2^{~2}+\bar{f}_1^{~2}f_2^{~1}
-(f_1^{~1}\bar{f}_2^{~2}+f_1^{~2}\bar{f}_2^{~1}) \\ 
\x^2&=& \bar{f}_{1}^{1}f^1_2-f^1_1\bar{f}^1_2 +
\bar{f}_{1}^{2}f^2_2-f^2_1\bar{f}^2_2 \\
\x^3&=& \bar{f}_{1}^{1}f^1_2-f^1_1\bar{f}^1_2
-\bar{f}_{1}^{2}f^2_2+f^2_1\bar{f}^2_2
\end{eqnarray}
while we have $\vec{\x}^2\equiv (\x^{1})^2 +(\x^{2})^2 + (\x^{3})^2  \neq 0$.
We now choose the direction $\x^{2}= \xi^{3}=0$ and $\xi^{1}=2i$, so that it
our constraints now take the form
\begin{eqnarray}
\bar{f}_{1}^{1}f_2^{~2}+\bar{f}_1^{~2}f_2^{~1}
-(f_1^{~1}\bar{f}_2^{~2}+f_1^{~2}\bar{f}_2^{~1})&=& 2i \\ 
(-(f_1^{~1})^{\ast}f_2^{~1}+f_1^{~1}(f_2^{~1})^{\ast})
+((f_1^{~2})^{\ast}f_2^{~2}-f_1^{~2}(f_2^{~2})^{\ast})&=& 2i
\label{stra1}
\end{eqnarray}
and 
\begin{equation}
f_2^{~1}(f_1^{~2})^{\ast}= f_1^{~1}(f_2^{~2})^{\ast}~~,\quad
f_2^{~2}(f_1^{~1})^{\ast}= f_1^{~2}(f_2^{~1})^{\ast} 
\label{stra2}
\end{equation}
We thus end up with only two+one real constraints and one gauge invariance
\begin{eqnarray}
\left( \begin{array}{c}f_1 \\f_2 \end{array}\right)'
= \left( \begin{array}{cc} \cos(\alpha) & \sin(\alpha)\\ -\sin(\alpha) &
\cos(\alpha) \end{array} \right)
\left( \begin{array}{c} f_1 \\ f_2\end{array} \right)
\end{eqnarray}
In the parametrization
\begin{eqnarray}
f_i^{~j}= m_i^{~j} exp(i \varphi_i^{~j})
\end{eqnarray}
the constraints (\ref{stra1}) and (\ref{stra2}) read
\begin{eqnarray}
m_1^1 m_2^2 = m_1^2m_2^1 
e^{-i\varphi_2^1-i\varphi_2^2+i\varphi_1^1+i\varphi_1^2} \\
m_1^1m_2^1\sin(\varphi_1^1-\varphi_2^1)+m_2^2m_1^2 
\sin(\varphi_2^1-\varphi_1^2)=1
\end{eqnarray}

We now want to fix the local $U(1)$ symmetry by imposing the gauge condition 
\begin{eqnarray}
\varphi_2^1+\varphi_2^2&=&\varphi_1^1+\varphi_1^2~.
\label{gauge}
\end{eqnarray}
When using
\begin{eqnarray}
\label{indep}
\abs{f_2^1}\equiv  m, \qquad \abs{f_2^2}\equiv n, \qquad \varphi_1^1 \equiv
 \theta~,\qquad \varphi_2^2 \equiv \phi~,
\end{eqnarray}
as the independent fields, our constraints above can be easily solved:
\begin{equation}
\label{varia}
-\varphi_2^1 = \varphi_2^2 = \phi, \qquad \varphi_1^1= -\varphi_1^2= \theta
\end{equation}
and
\begin{eqnarray}
m_1^1 = \fracmm{m}{(m^2+n^2)\sin(\theta+\phi)}~,\quad 
m_1^2 = \fracmm{n}{(m^2+n^2)\sin(\theta+\phi)}~,\quad
m_2^1=m, \quad m_2^2 = n
\end{eqnarray}

It is straightforward to deduce the other fields $F_i^+, A_{im}^-, M_i^-$ and 
$N_i^-$ in terms of the independent components (\ref{indep}). The scalar 
potential (\ref{pot}) in terms of these independent field variables 
takes the form (no indices and constraints any more~!)
\begin{equation}
V= \fracmm{\abs{Z}^2\sin^2(\theta +\phi)}{m^2+n^2}\left[
\fracmm{4(m^2-n^2)^2}{1+(m^2+n^2)^2\sin^2(\theta +\phi)}
+\fracmm{1+(m^2+n^2)^2\sin^2(\theta +\phi)}{\sin^4(\theta +\phi)}\right]
\end{equation}
It is clear from this equation that the potential $V$ is positively definite, 
and it is only non-vanishing due to the non-vanishing central charge $\abs{Z}$.
It signals the spontaneous breaking of $N=2$ supersymmetry in our model.
\vglue.2in


\begin{thebibliography}{9}
\bibitem{all} S. V. Ketov, {\it On the exact solutions to N=2 supersymmetric
gauge theories}, DESY and Hannover preprint, DESY 97--199 and ITP-UH-26/97,
October 1997; hep-th/9710085.
\bibitem{ikz} E. A. Ivanov, S. V. Ketov and B. M. Zupnik, {\it Induced 
hypermultiplet self-interactions in N=2 gauge theories}, DESY, Dubna and
Hannover preprint, DESY 97--094, JINR E2--97--164 and ITP--UH--10/97, June
1997; hep-th/9706078; to appear in Nucl. Phys. B.
\bibitem{gios} A. A. Galperin, E. A. Ivanov, V. I. Ogievetsky and E. Sokatchev,
\cmp{103}{86}{515}.
\bibitem{witten} E. Witten, {\it Solutions of four-dimensional field theory via
M theory}, Princeton preprint IASSNS--HEP--97--19 March 1997; hep-th/9703166. 
\bibitem{giot} A. A. Galperin, E. A. Ivanov, V. I. Ogievetsky 
and P. K. Townsend, \cqg{3}{86}{625}.
\end{thebibliography}
\end{document}